\documentclass[aps,floatfix,prb,twocolumn,showpacs]{revtex4}%
\usepackage{amsmath}
\usepackage[final,dvips]{graphicx}
\usepackage{amsfonts}
\usepackage{amssymb}%
\providecommand{\U}[1]{\protect\rule{.1in}{.1in}}
\begin{document}
\title{The effect of Magnetic Field on MQT Escape Time \\in ``Small" Josephson Junctions}
\author{Yu.N.Ovchinnikov}
\affiliation{Max-Plank Institute for Physics of Complex Systems, Dresden, D-01187 Germany
and Landau Institute for Theoretical Physics, RAS, Chernogolovka, Moscow
District, 142432 Russia}
\author{A.Barone}
\affiliation{Dipartimento Scienze Fisiche, Universit\`{a} di Napoli \textquotedblleft
Federico II" and Coherentia-INFM, CNR, piazzale Tecchio, 80, 80125, Napoli, Italy}
\author{A.A.Varlamov}
\affiliation{Coherentia-INFM, CNR, Viale del Politecnico 1, I-00133 Rome, Italy }
\date{\today}

\begin{abstract}
We study the phenomenon of macroscopic quantum tunneling (MQT) in a finite
size Josephson junction (JJ) with an externally applied magnetic field. As it
is well known, the problem of MQT in a point-like JJ is reduced to the study
of the under-barrier motion of a quantum particle in the washboard potential.
In the case of a finite size JJ placed in magnetic field, this problem is
considerably more complex since, besides the phase, the potential itself,
depends on space variables. We find the general expressions both for the
crossover temperature $T_{0}$ between thermally activated and macroscopic
quantum tunneling regimes and the escape time $\tau_{\mathrm{esc}}$. It turns
out that in proximity of particular values of magnetic field the crossover
temperature can vary non-monotonically.

\end{abstract}

\pacs{74.50.+r}
\maketitle

\section{Introduction}

The Josephson effect \cite{J62,BP82} allows the investigation of fundamental
aspects of quantum phenomena such as the macroscopic quantum tunneling
(MQT)\cite{L80} which has been more recently observed for the first time also
in high-$T_{c}$ bi-epitaxial $YBCO$ junctions\cite{B2005,I2005}. Interesting
results have been obtained also in various $Bi-2212$ structures particularly
referred to ``intrinsic'' Josephson junctions.

Usually the phenomenon of macroscopic quantum tunneling is considered in a
``point''- like Josephson Junctions (JJ), i.e. completely neglecting the
finiteness of the junction size $L$ (some exceptions can be found in
theoretical and experimental \cite{a} papers \cite{b,ba}). This, zero order in
$L,$ approximation is based on the assumption that the junction size is much
smaller than all other related parameters of the problem, such as the
Josephson penetration depth $\lambda_{J}=\left(  \hbar c^{2}/8\pi ej_{c}%
\ell_{\mathrm{eff}}\right)  ^{1/2}$ and the characteristic length $L_{H}%
=\ell_{H}^{2}/\ell_{\mathrm{eff}}$, (with $\ell_{H}=\left(  \hbar c/eH\right)
^{1/2}$ as the standard quantum mechanical magnetic length)\cite{c}. The
effective length $\ell_{\mathrm{eff}}$ depends on the relation between the
thickness $d_{\left(  i\right)  }$ $(i=L,R)$ of the superconductive electrodes
and the London penetration depth $\lambda_{\left(  i\right)  }$ of the bulk
superconductor materials, which the electrodes are made of. In the limiting
cases one can find \cite{EST64}
\begin{equation}
\ell_{\mathrm{eff}}=\left\{
\begin{tabular}
[c]{ll}%
$\lambda_{\left(  1\right)  }+\lambda_{\left(  2\right)  }+d_{\mathrm{ox}},$ &
$\lambda_{\left(  i\right)  }\ll d_{\left(  i\right)  }$\\
$d_{\left(  1\right)  }+d_{\left(  2\right)  }+d_{\mathrm{ox}},$ & $d_{\left(
i\right)  }\ll\lambda_{\left(  i\right)  }$%
\end{tabular}
\ \ \ \ \ \ \ \ \ \ \ \ \ \ \ \right.  . \label{leff}%
\end{equation}

Thermal fluctuations in Josephson junctions,\ produce a typical "rounding" of
the Josephson current branch in the I-V curves \cite{AH69,Hermann2008}. Since
the pioneering measurements of \ thermal fluctuation phenomena, to obtain such
an effect, the required condition between thermal energy \ and Josephson
coupling energy ($E_{J}\sim T)$, was usually realized not by increasing the
temperature, rather, by reducing the value of $E_{J}$ by applying a proper
magnetic field. This, even in the case of small JJ ($L<\lambda_{J}$), has
significant implications which become of paramount importance in MQT activation.

In this context the authors \cite{OBV07} reported the analysis of the role of
finiteness of the junction's length $L$ obtaining the general expression for
the crossover temperature $T_{0}$ between thermally activated and MQT regimes
for such JJ. The escape time $\tau_{\mathrm{esc}}$ was calculated with the
exponential accuracy in the first approximation in $\left(  L/L_{H}\right)
^{2}$ for temperatures $T>T_{0}$ and for small region below $T_{0}.$ It was
demonstrated that the account for the junction's size results in the
appearance of a strong sensitivity of the MQT process on applied magnetic
field, making the crossover temperature to be non-monotonic function of it.
Since magnetic field is an easily adjustable parameter, it can become an
important tool in study of such a quantum coherent phenomenon without
modification of other junction parameters.

In the present article we will proceed and develope the study of MQT in a
finite size Josephson junction (JJ) placed in magnetic field. First we will
report the mean field solutions for the effective action of such a finite size
Josephson system and find the values of the effective action at the extremal
trajectories. Then, the explicit form of phase trajectories close to the
extremal ones and corresponding action functional will be found. This will
allow us to find the pre-exponential factor in the expression for
$\tau_{\mathrm{esc}}$ of such extended system in a wide temperature region,
including the crossover point $T_{0}$. It turns out that in the vicinity of
magnetic fields values $H_{n}=\Phi_{0}n/\left(  L\ell_{\mathrm{eff}}\right)  $
($\Phi_{0}=\pi\hbar c/e$ is the magnetic flux quantum, $c$ is light velocity),
the escape time can vary non-monotonically, a new phenomenon which becomes the
fingerprint for the experimental check of the proposed theory.

In our analysis we will suppose that $L\ll\lambda_{J}$ but will not impose any
restrictions on the relation between $L$ and $L_{H}.$

\section{Generalities}

\subsection{Escape time}

Starting the discussion of the phenomenon of Josephson current decay in a
finite size Josephson junction let us recall the substance of this process in
a point-like junction.

Let us consider a current biased Josephson tunnel junction. In the framework
of the capacitively shunted junction model it can be represented by the
electronic equivalent circuit, where the resistance of the external circuit
($R_{\mathrm{ext}})$, the intrinsic junction capacitance ($C$) and junction
resistance ($R_{T}$), assumed as a linear ohmic element, are connected in
parallel \cite{BP82}. The current balance in the circuit can be accounted by
the following equation
\begin{equation}
I=I_{\mathrm{c}}\sin2{\phi}+\frac{V}{R_{T}}+C\frac{dV}{dt},\label{Josephsoneq}%
\end{equation}
where
\begin{equation}
\phi=e\int Vdt\label{phiint}%
\end{equation}
is the relative phase between the two superconductors. Equation
(\ref{Josephsoneq}) can be rewritten in the form:
\begin{equation}
M_{C}\frac{\partial^{2}\phi}{\partial t^{2}}+\eta\frac{\partial\phi}{\partial
t}+\frac{\partial U\left(  \phi\right)  }{\partial\phi}=0,\label{eqmot}%
\end{equation}
where $M_{C}=\hbar^{2}C/e^{2}$, $\eta=\hbar\left(  e^{2}R\right)  ^{-1}$
\begin{equation}
U(\phi)=-(\frac{\hbar I}{e}\phi+E_{J}\cos{2\phi}),\qquad E_{J}=\frac{\hbar
I_{\mathrm{c}}}{2e}.\label{washeq}%
\end{equation}
\begin{figure}[ptb]
\centerline{\includegraphics[width=.3\textwidth]{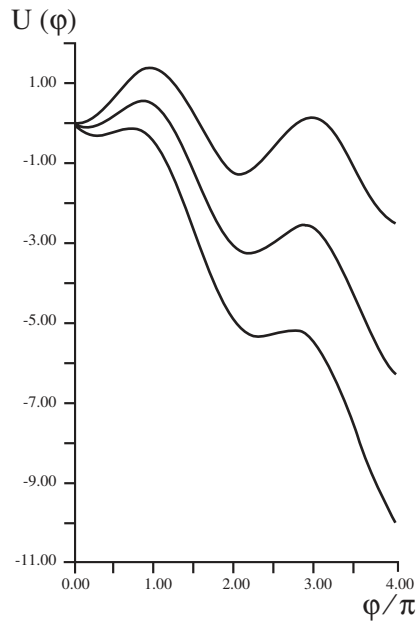}}\caption{
Washboard potential.}%
\label{washb}%
\end{figure}Eq. (\ref{eqmot}) can be considered as the equation of motion with
friction ($\eta$ is its viscosity) of a particle of mass $M_{C}$ in the
washboard potential (\ref{washeq}) (see Fig. \ref{washb}). The values of the
bias current $I$ and the critical current $I_{\mathrm{c}}$ determine the slope
of the potential $U(\phi)$ and the depth of valleys.

Let us study the case $I<I_{\mathrm{c}}$. The minima of potential
(\ref{washeq}) correspond to the metastable states of the junction. We will
assume the junction viscosity to be small enough not to affect noticeably the
particle oscillations in the well and its under-barrier motion. At high
temperatures the thermally activated escape dominates and the result of
classical Kramers problem of a particle moving in the washboard potential
$U\left(  x\right)  $ is valid:
\begin{equation}
\Gamma=\left[  \tau_{\mathrm{esc}}^{\left(  \mathrm{th}\right)  }\right]
^{-1}\approx\frac{\omega_{\mathrm{p}}}{2\pi}\exp{\left[  -\frac{\Delta U}%
{T}\right]  ,} \label{arren}%
\end{equation}
where the characteristic plasma frequency $\omega_{\mathrm{p}}$ is given by
the interpolation formula
\begin{equation}
\omega_{\mathrm{p}}=\left[  \frac{1}{M_{C}}\frac{d^{2}U(\phi)}{d\phi^{2}%
}\right]  _{\phi_{\mathrm{min}}}^{\frac{1}{2}}=\left(  \frac{2\pi
cI_{\mathrm{c}}}{\Phi_{0}C}\right)  ^{1/2}\left[  1-\left(  \frac
{I}{I_{\mathrm{c}}}\right)  ^{2}\right]  ^{1/4} \label{plasma}%
\end{equation}
and $\Delta U=U_{\max}-U_{\min}$. The extremes of potential $U_{\max}$ and
$U_{\min}$ correspond to the local maximum and local minimum of the potential energy.

When the temperature decreases thermal activation becomes less and less
probable and the process of quantum tunneling through the barrier becomes
important. Near some temperature $T_{0}$ the crossover between Arrenius law
and quantum tunneling regime takes place. The latter can be considered as the
underbarrier motion of the particle (instanton propagation).

At very low temperatures $\left(  T\ll\omega_{\mathrm{p}}\right)  $ the
tunneling takes place only from the lowest level of the energy spectrum. When
$I_{\mathrm{c}}-I\ll I_{\mathrm{c}}$ the escape rate can be presented in the
form \cite{who??}:%

\begin{equation}
\Gamma=\left[  \tau_{\mathrm{esc}}^{\left(  \mathrm{qm}\right)  }\right]
^{-1}=\frac{6\omega_{\mathrm{p}}}{\sqrt{\pi}}\left(  \frac{6\Delta U}%
{\omega_{\mathrm{p}}}\right)  ^{1/2}\exp\left[  -\frac{36\Delta U}%
{5\omega_{\mathrm{p}}}\right]  . \label{prob2}%
\end{equation}
One can see that the main difference between Eqs. (\ref{arren}) and
(\ref{prob2}) consists only in the temperature dependence of the exponent,
whereas in the classical case this is the activation Boltzmann exponential law
with the exponent equal to the barrier height divided $T$, in the case of the
quantum tunneling temperature in exponent has to be substituted by
$\omega_{\mathrm{p}}$.

At an arbitrary temperature the combined tunneling occurs by the following
scheme. First, the particle excites in the thermal activation manner and at
some moment gets up to the energy level $E_{n}$ (calculated neglecting
tunneling) and then by means of quantum tunneling passes through the barrier.
The total tunneling probability is determined by the sum over all quantum
levels of corresponding products of the classical and quantum probabilities:
\begin{equation}
\Gamma\sim\sum_{n}\exp\left(  -\frac{E_{n}}{T}\right)  \exp\left\{  -A\left(
E_{n}\right)  \right\}  . \label{qm}%
\end{equation}
Here%
\begin{equation}
A\left(  E_{n}\right)  =%
{\displaystyle\oint}
\sqrt{2M_{C}\left[  E_{n}-U(\phi)\right]  }d\phi\label{Smelnik}%
\end{equation}
is the action corresponding to the under-barrier motion of the particle with
energy $E_{n}.$

One can say that quantum tunneling occurs when the particle reaches such a
level $E_{\mathrm{tun}}$ at which the probability of the direct quantum
tunneling through the barrier becomes larger than the probability of the
activation jump on the higher energy levels with further tunneling through the
barrier. Quantitatively this can be formulated as the condition for the
extremum of the exponent in Eq. (\ref{qm})%

\begin{equation}
\left.  \frac{\partial A\left(  E\right)  }{\partial E}\right\vert
_{E_{\mathrm{tun}}}=-\frac{1}{T}. \label{extre}%
\end{equation}
Condition (\ref{extre}) implies that the period of the particle oscillatoin in
the inverted potential is equal to $1/T$. With the growth of temperature the
energy $E_{\mathrm{tun}}$ increases and when $T=\omega_{\mathrm{p}}/2\pi$ it
reaches the barrier height. At higher temperatures the classical activation
escape scheme is realized.

The value of the escape time $\tau_{\mathrm{esc}}$ of the finite size JJ in
magnetic field can be determined in the framework of the general method of
functional integration. In this approach the escape rate of the MQT, which is
defined by the imaginary part of the free energy $\tau_{\mathrm{esc}}%
^{-1}=2\operatorname{Im}\emph{F,}$ can be expressed in terms of the partition
function of the system \cite{LO84}:
\begin{equation}
\emph{F}=-T\ln Z, \label{freeenergy}%
\end{equation}
where the latter is defined by the functional integral of the S-matrix over
all possible trajectories $\varphi\left(  \mathbf{r},\tau\right)  :$
\begin{equation}
Z=\int\emph{D}\varphi\left(  \mathbf{r},\tau\right)  \exp\left(  -A\left[
\varphi\left(  \mathbf{r},\tau\right)  \right]  \right)  . \label{PF}%
\end{equation}
Here $\varphi\left(  \mathbf{r},\tau\right)  $ is the phase difference on the
junction at the point $\mathbf{r}$\ and imaginary time $\tau$ and $A\left[
\varphi\left(  \mathbf{r},\tau\right)  \right]  $ is the effective action of
such extended system.

The problem of MQT makes sense only within the framework of the
quasi-classical approximation, i.e. to assume that the value of escape time
considerably exceeds all characteristic time-scales of the internal motions.
In this case, the imaginary part of the partition function is small in
comparison with the real one, hence%

\begin{equation}
\tau_{\mathrm{esc}}^{-1}=2T\frac{\operatorname{Im}\int\emph{D}\varphi\left(
\mathbf{r},\tau\right)  \exp\left(  -A\left[  \varphi\left(  \mathbf{r}%
,\tau\right)  \right]  \right)  }{\operatorname{Re}\int\emph{D}\varphi\left(
\mathbf{r},\tau\right)  \exp\left(  -A\left[  \varphi\left(  \mathbf{r}%
,\tau\right)  \right]  \right)  }. \label{escape}%
\end{equation}

The imaginary part in the numerator of Eq. (\ref{escape}) is determined by the
saddle trajectory of action $\varphi^{\mathrm{sdl}}\left(  \mathbf{r}%
,\tau\right)  ,$ while the real part of the partition function in denominator
has to be calculated at the minimal trajectory $\varphi^{\mathrm{\min}}\left(
\mathbf{r},\tau\right)  .$ For high enough temperatures $T>T_{0}$, or in
narrow vicinity of $T_{0}$ ( $|T-T_{0}|\ll T_{0}$), both trajectories
$\varphi^{\mathrm{sdl}}\left(  \mathbf{r}\right)  $ and $\varphi^{\min}\left(
\mathbf{r}\right)  $ turn out to be time independent \cite{LO84,OS93} and the
exponential factor in $\tau_{\mathrm{esc}}^{-1},$ in analogy with Eq.
(\ref{arren}), is determined by the expression
\begin{align}
\tau_{\mathrm{esc}}^{-1}  &  \sim\exp\left[  -\Delta A_{\min}^{\mathrm{sdl}%
}\right]  ,\qquad\nonumber\\
\Delta A_{\min}^{\mathrm{sdl}}  &  =A\left[  \varphi^{\mathrm{sdl}}\left(
\mathbf{r}\right)  \right]  -A\left[  \varphi^{\min}\left(  \mathbf{r}\right)
\right]  . \label{escdeltaA}%
\end{align}
\

In the assumption of zero viscosity,\ one can obtain both thermally activation
and MQT escaped times (the latter even with pre-exponential factor accuracy)
\cite{LO84}. \ For temperatures above but not too close to $T_{0}$ it is read
as
\begin{equation}
\tau_{\mathrm{esc}}^{-1}=T_{0}\frac{\sinh\left(  \frac{\omega_{\mathrm{p}}%
}{2T}\right)  }{\sin\left(  \pi T_{0}/T\right)  }\exp\left\{  -\Delta A_{\min
}^{\mathrm{sdl}}\right\}  . \label{escpre}%
\end{equation}

Hence, in order to find the escape time\ (Eqs. (\ref{escdeltaA})-(\ref{escpre}%
)) one has to calculate the values of action for trajectories $\varphi
^{\mathrm{sdl}}$ and $\varphi^{\min}$. The crossover temperature $T_{0}$ turns
to be the bifurcation point, below which the time dependent solution for the
saddle point trajectory $\varphi^{\mathrm{sdl}}\left(  \mathbf{r},\tau\right)
$ becomes more favorable than the static one and the definition of escape time
presents more sophisticated problems.

\subsection{Effective action}

The effective action of a Josephson junction placed in external magnetic field
$H$ as a function of flowing through the junction current $I$, can be written
down in the most general case basing on the results of \cite{LO83a,LO83b}:
\begin{widetext}
\begin{align}
A\left[  \varphi\left(  \mathbf{r},\tau\right)  \right]  =  &  \frac{1}{S}%
\int_{-1/2T}^{1/2T}d\tau\int d^{2}\mathbf{r}\left\{  \frac{C}{2e^{2}}\left(
\frac{\partial\varphi\left(  \mathbf{r},\tau\right)  }{\partial\tau}\right)
^{2}-\frac{\hbar}{e}I\varphi\left(  \mathbf{r},\tau\right)  \right.
-\frac{\pi\hbar}{2R_{\mathrm{N}}e^{2}}\int_{-1/2T}^{1/2T}d\tau_{1}\left\{
\left\{  1-\cos\left[  \varphi\left(  \mathbf{r},\tau\right)  -\varphi\left(
\mathbf{r},\tau_{1}\right)  \right]  \right\}  \cdot\right. \nonumber\\
&  \left.  \alpha_{L}\left(  \tau-\tau_{1}\right)  \alpha_{R}\left(  \tau
-\tau_{1}\right)  +\cos\left[  \varphi\left(  \mathbf{r},\tau\right)
+\varphi\left(  \mathbf{r},\tau_{1}\right)  \right]  \beta_{L}\left(
\tau-\tau_{1}\right)  \beta_{R}\left(  \tau-\tau_{1}\right)  \right\}
+\frac{\pi\hbar T^{2}}{R_{\mathrm{sh}}e^{2}}\cdot\nonumber\\
&  \left.  \int_{-1/2T}^{1/2T}d\tau_{1}\frac{\sin^{2}\left(  \left[
\varphi\left(  \mathbf{r},\tau\right)  -\varphi\left(  \mathbf{r},\tau
_{1}\right)  \right]  /2\right)  }{\sin^{2}\left[  \pi T\left(  \tau-\tau
_{1}\right)  \right]  }\right\}  +\frac{\hbar^{2}c^{2}}{8\pi e^{2}%
\ell_{\mathrm{eff}}}\int_{-1/2T}^{1/2T}d\tau\int d^{2}\mathbf{r}\left(
\frac{\partial\varphi\left(  \mathbf{r},\tau\right)  }{\partial\mathbf{r}%
}-\frac{e\ell_{\mathrm{eff}}}{\hbar c}\left[  \mathbf{H}\times\mathbf{n}%
\right]  \right)  ^{2}. \label{Aphi}%
\end{align}
\end{widetext}Here $C$ is the junction capacitance, $R_{\mathrm{N}}$ is the
tunnel resistance, $R_{\mathrm{sh}}$\ is the shunt resistance, $\alpha
_{\left(  L,R\right)  }\left(  \tau\right)  $ and $\beta_{\left(  L,R\right)
}\left(  \tau\right)  $ are integrated over the energy variable normal and
anomalous Green functions. The integrals are carried out over the imaginary
time and the junction area. The first term corresponds to the kinetic energy
of the junction. The second and fourth terms describe the contribution of the
potential energy to the action. Let us stress that the third term,
corresponding to the capacitance renormalization, appears only beyond the
quasi-classical \ approximation. The last term in Eq. (\ref{Aphi}) accounts
for the finite size of the junction and the magnetic field contribution to the
effective action.\cite{class}

The natural assumption $T_{0}\ll T_{c}$ side by side with the assumed above
condition $L\ll\lambda_{J}$ allows to considerably simplify the general
expression Eq. (\ref{Aphi}), what was done in the leading approximation in
$\left(  L/\lambda_{J}\right)  ^{2}$ in Ref. \cite{OBV07}. It was shown that
in these conditions the third term in Eq. (\ref{Aphi}) disappears being
reduced to renormalization of the capacity $C$ in the first term
\begin{equation}
C^{\ast}=C+\frac{\pi\hbar}{2R_{\mathrm{N}}}\int_{-\infty}^{\infty}%
\frac{d\omega}{2\pi}\left(  \frac{\partial\alpha_{L}}{\partial\omega}\right)
\left(  \frac{\partial\alpha_{R}}{\partial\omega}\right)
\end{equation}
and changing in the definite way the shape of $\varphi\left(  \mathbf{r}%
,\tau\right)  $.

Variation of \ Eq. (\ref{Aphi}) on $\varphi$ results in getting of the
quasi-classical equations of motion \cite{EST64}, which define the extremal
trajectories%
\begin{equation}
\left.  \frac{\delta A\left[  \varphi\right]  }{\delta\varphi}\right\vert
_{\varphi=\varphi^{\mathrm{extr}}\left(  \mathbf{r},\tau\right)  }=0.
\label{af0}%
\end{equation}
Near the extremal trajectory $\varphi^{\mathrm{extr}}\left(  \mathbf{r}%
,\tau\right)  $ the deviation of the function $\varphi\left(  \mathbf{r}%
,\tau\right)  $ can be represented in the form of expansion by normalized
functions $\varphi_{n}^{k}\left(  \mathbf{r},\tau\right)  $
\begin{equation}
\varphi\left(  \mathbf{r},\tau\right)  =\varphi^{\mathrm{extr}}\left(
\mathbf{r},\tau\right)  +\sum B_{n}^{k}\varphi_{n}^{k}\left(  \mathbf{r}%
,\tau\right)  \label{expan}%
\end{equation}
which are the eigenfunctions of the equation%
\begin{equation}
\frac{\delta^{2}A\left[  \varphi\right]  }{\delta\varphi^{2}}\varphi_{n}%
^{k}\left(  \mathbf{r},\tau\right)  =\Lambda_{n}^{k}\varphi_{n}^{k}\left(
\mathbf{r},\tau\right)  . \label{expan1}%
\end{equation}
In this representation the action (\ref{Aphi}) near the extremal trajectory
$\varphi^{\mathrm{extr}}\left(  \mathbf{r},\tau\right)  $ is presented by
Gaussian type functional integral over functions $\varphi_{n}^{k}\left(
\mathbf{r},\tau\right)  $ which can be carried out analytically. Thus, the
problem of definition of the value of escape time is reduced to finding of the
eigenvalues $\Lambda_{n}^{k}$ of the Eq. (\ref{expan1}) \cite{LO84}.

We will restrict our consideration within two regions of temperatures: a).
$T>T_{0}$ and b) $|T-T_{0}|\ll$ $T_{0}.$ This choice is related to the fact
that in both these regions one can use the functions $\varphi\left(
\mathbf{r},\tau\right)  $ in the form Eq. (\ref{expan}) with time independent
$\varphi^{\mathrm{extr}}\left(  \mathbf{r}\right)  .$ The width of the
crossover region between thermal activation and macroscopic quantum tunneling
regimes\ turns to be much smaller than $T_{0}$ and will be estimated below.

\section{Extremal trajectories in static approximation}

Let us find the explicit form of the extremal trajectory $\varphi
^{\mathrm{extr}}$ in static approximation, i.e. in the case when it can be
considered as time independent. \ Moreover, in the geometry under
consideration, when the magnetic field is applied along the junction, the
phase depends only on one coordinate $x$. Substitution of $\varphi
^{\mathrm{extr}}\left(  x,\tau\right)  \equiv\varphi^{\mathrm{extr}}\left(
x\right)  $ in Eq. ( \ref{af0}) leads to the equation
\begin{equation}
-\lambda_{J}^{2}\frac{\partial^{2}\varphi^{\mathrm{extr}}\left(  x\right)
}{\partial x^{2}}+\frac{1}{2}\sin2\varphi^{\mathrm{extr}}\left(  x\right)
=\frac{I}{2j_{c}S}. \label{extremum}%
\end{equation}
We will look for its solutions in the form
\begin{equation}
\varphi^{\mathrm{extr}}\left(  x\right)  =\varphi_{0}-\frac{x}{L_{H}%
}+\widetilde{\varphi}\left(  x\right)  , \label{phitot}%
\end{equation}
where $\varphi_{0}$ is \ a constant, and $\left\langle \widetilde{\varphi
}\left(  x\right)  \right\rangle =0.$ Corresponding boundary conditions are
\begin{equation}
\left.  \frac{\partial\widetilde{\varphi}}{\partial x}\right\vert
_{x=-L/2}=\left.  \frac{\partial\widetilde{\varphi}}{\partial x}\right\vert
_{x=L/2}=0. \label{bound}%
\end{equation}

One can look for solutions of \ Eq. (\ref{extremum}) in the frameworks of the
perturbation theory by the parameter $\left(  L/\lambda_{J}\right)  ^{2}.$ In
the first approximation one can rewrite Eq. (\ref{extremum}) in the form%
\begin{equation}
-\lambda_{J}^{2}\frac{\partial^{2}\widetilde{\varphi}\left(  x\right)
}{\partial x^{2}}=\frac{I}{2Sj_{c}}-\frac{1}{2}\sin\left[  2\left(
\varphi_{0}-\frac{x}{L_{H}}\right)  \right]  . \label{perturb1}%
\end{equation}
This equation with the given above boundary conditions is easily solvable:%

\begin{align}
\widetilde{\varphi}\left(  x\right)   &  =\frac{1}{2}\left(  \frac{L_{H}%
}{2\lambda_{J}}\right)  ^{2}\left\{  \left[  \sin\left(  \frac{2x}{L_{H}%
}\right)  -\frac{2x}{L_{H}}\cos\frac{L}{L_{H}}\right]  \cos2\varphi_{0}\right.
\nonumber\\
&  +\left[  \frac{L_{H}}{L}\sin\left(  \frac{L}{L_{H}}\right)  -\cos\left(
\frac{2x}{L_{H}}\right)  \right. \nonumber\\
&  \left.  \left.  +\sin\left(  \frac{L}{L_{H}}\right)  \left(  \frac
{L}{6L_{H}}-\frac{2x^{2}}{LL_{H}}\right)  \right]  \sin2\varphi_{0}\right\}  .
\label{phigen}%
\end{align}

The phase $\varphi\left(  x,\varphi_{0}\right)  $ is a periodic function of
$\varphi_{0}$ with the period $\pi.$ It turns out that some critical value of
the current $I_{\mathrm{cr}}\left(  H\right)  $ exists such, that for all
\ $I<I_{\mathrm{cr}}$ one can find for each period two solutions for
$\varphi_{0}$. The exception presents narrow regions of certain magnetic field
values, which will be found and investigated below. It will be shown that for
such regions four different solutions for $\varphi_{0}$ exist in the interval
]$-\pi/2,\pi/2]$. Two solutions correspond to the minimum of action while the
other two to its saddle point. When $I=I_{\mathrm{cr}}$ pair solutions (one
corresponding to the minimum, the other to the saddle point of the action)
confluent in one, while for $I>I_{\mathrm{cr}}$ the static solution does not
exist at all. In the simple case of a point-like junction one can find:%
\begin{equation}
\varphi_{00}^{\min}=\frac{1}{2}\arcsin\left(  \frac{I}{j_{c}S}\right)
\label{phimin}%
\end{equation}
and%
\begin{equation}
\varphi_{00}^{\mathrm{sdl}}=\frac{\pi}{2}-\frac{1}{2}\arcsin\left(  \frac
{I}{j_{c}S}\right)  . \label{phisaddle}%
\end{equation}

When the junction has finite width the analysis is more complicated.
Nevertheless the smallness $L\ll\lambda_{J}$ allows us to find the relation
between $\varphi_{0}$ and $I,$ i.e. to find $\varphi^{\mathrm{extr}}\left(
x,I\right)  $. Indeed, integrating Eq. (\ref{extremum}) \ over $x$, one can
obtain:
\begin{equation}
I=j_{c}S\int_{-L/2}^{L/2}\frac{dx}{L}\sin\left(  2\varphi_{0}-\frac{2x}{L_{H}%
}+2\widetilde{\varphi}\left(  x\right)  \right)  . \label{iphi}%
\end{equation}
Substituting in this expression $\widetilde{\varphi}\left(  x\right)  $
according to Eq. (\ref{phigen}) and performing integration one finds the
equation which implicitly relates $\varphi_{0}$ to $I:$
\begin{align}
\frac{L_{H}}{L}\sin\left(  \frac{L}{L_{H}}\right)  \sin2\varphi_{0}+\frac
{1}{2}\left(  \frac{L_{H}}{2\lambda_{J}}\right)  ^{2}\cdot\nonumber\\
\kappa\left(  \frac{L_{H}}{L}\right)  \sin4\varphi_{0} =\frac{I}{j_{c}S},
\label{iphitot}%
\end{align}
where%
\begin{align}
\kappa\left(  \frac{L_{H}}{L}\right)   &  =2\left(  \frac{L_{H}}{L}\right)
^{2}\sin^{2}\left(  \frac{L}{L_{H}}\right)  +\cos^{2}\left(  \frac{L}{L_{H}%
}\right) \nonumber\\
&  -\frac{3L_{H}}{2L}\sin\left(  \frac{2L}{L_{H}}\right)  -\frac{1}{3}\sin
^{2}\left(  \frac{L}{L_{H}}\right)  . \label{kappa}%
\end{align}
The critical current $I_{\mathrm{cr}}\left(  H\right)  $ is determined by the
value $I\left(  \widehat{\varphi}_{0}\right)  ,$ where the point
$\widehat{\varphi}_{0}$ is solution of the equation%
\begin{equation}
\left.  \frac{\partial I\left(  \varphi_{0}\right)  }{\partial\varphi_{0}%
}\right\vert _{\varphi_{0}=\widehat{\varphi}_{0}}=0,\qquad I_{\mathrm{cr}%
}\left(  H\right)  =I\left(  \widehat{\varphi}_{0}\right)  . \label{icr}%
\end{equation}

\section{The value of effective action at the extremal trajectories}

Let us find the value of effective action of the finite size JJ on the
extremal trajectory ${\varphi^{\mathrm{extr}}}\left(  x\right)  $. The
knowledge of $A\left[  \varphi^{\mathrm{extr}}\left(  x\right)  \right]  $
will allow us to find the value of escape time with the exponential accuracy
in the wide interval of temperatures above $T_{0},$ or close enough to this
bifurcation point ($|T-T_{0}|\ll T_{0}$ \cite{LO84}). Further definition of
the pre-exponential factor will require to perform the functional integration
in Eq. (\ref{escape}) over the trajectories close to the extremal one.

In the case when the phase trajectory $\varphi\left(  x,\tau\right)
\equiv\varphi\left(  x\right)  $ does not depend on imaginary time (static
approximation) Eq. (\ref{Aphi}) is considerably simplified:
\begin{align}
A\left[  \varphi\left(  x\right)  \right]   &  =\frac{\hbar Sj_{c}}{eLT}%
\int_{-L/2}^{L/2}dx\left\{  -\frac{I}{Sj_{c}}\varphi\left(  x\right)  \right.
\nonumber\\
&  \left.  -\frac{1}{2}\cos2\varphi\left(  x\right)  +\lambda_{J}^{2}\left(
\frac{\partial\varphi\left(  x\right)  }{\partial x}+\frac{1}{L_{H}}\right)
^{2}\right\}  . \label{aphistat}%
\end{align}
The straightforward integration of this expression with the phase ${\varphi
}\left(  x\right)  $ determined from the Eqs. ( \ref{phitot}), ( \ref{phigen})
results in%

\begin{align}
A\left(  \varphi_{0}\right)   &  =-\frac{\hbar I}{eT}\varphi_{0}-\frac{\hbar
Sj_{c}}{2eT}\left(  \frac{L_{H}}{L}\right)  \sin\left(  \frac{L}{L_{H}%
}\right)  \cos2\varphi_{0}\nonumber\\
&  +\frac{\hbar Sj_{c}}{8eT}\left(  \frac{L_{H}}{2\lambda_{J}}\right)
^{2}\left[  \mu\left(  \frac{L_{H}}{L}\right)  -\kappa\left(  \frac{L_{H}}%
{L}\right)  \cos4\varphi_{0}\right]  \label{aphi0}%
\end{align}
with%
\begin{equation}
\mu\left(  \frac{L_{H}}{L}\right)  =2\left\{  \left[  \left(  \frac{L_{H}}%
{L}\right)  ^{2}+\frac{1}{3}\right]  \sin^{2}\left(  \frac{L}{L_{H}}\right)
-1.\right\}  . \label{mu}%
\end{equation}

This expression already gives in explicit form the value of action both for
minimal and saddle trajectories (determined as function of current and
magnetic field by Eq. (\ref{iphitot})). Let us mention that Eq. (\ref{iphitot}%
) could be obtained from Eq. (\ref{aphi0}) deriving the action and equating
this derivative with zero: $\partial A\left(  \varphi_{0}\right)
/\partial\varphi_{0}=0.$ Moreover, calcuating the second derivative of the
action (\ref{aphi0}) one find another useful relation%
\begin{align}
\frac{\partial^{2}A\left(  \varphi_{0}\right)  }{\partial^{2}\varphi_{0}}  &
=\frac{2\hbar Sj_{c}}{eT}\left[  \left(  \frac{L_{H}}{L}\right)  \sin\left(
\frac{L}{L_{H}}\right)  \cos2\varphi_{0}\right. \nonumber\\
&  +\left.  \left(  \frac{L_{H}}{2\lambda_{J}}\right)  ^{2}\kappa\left(
\frac{L_{H}}{L}\right)  \cos4\varphi_{0}\right]  . \label{20**}%
\end{align}

Note that both Eqs. (\ref{iphitot}) and (\ref{aphi0}) are valid for any value
of magnetic field, even in the region where the second (correction) term in
Eq. (\ref{iphitot}) becomes of the order or even larger than the first one.
Eq. (\ref{icr}) for the value of $\widehat{\varphi}_{0}$ can be written in the
form:%
\begin{equation}
\delta_{1}\cos2\widehat{\varphi}_{0}+2\delta_{2}\cos4\widehat{\varphi}_{0}=0,
\label{sico}%
\end{equation}
where
\begin{equation}
\delta_{1}=\frac{L_{H}}{L}\sin\left(  \frac{L}{L_{H}}\right)  \text{ \ and
}\delta_{2}=\frac{1}{2}\left(  \frac{L_{H}}{2\lambda_{J}}\right)  ^{2}%
\kappa\left(  \frac{L_{H}}{L}\right)  . \label{delty}%
\end{equation}
Solutions of Eq. (\ref{sico}) are%
\begin{equation}
\cos2\widehat{\varphi}_{0}=-\frac{\delta_{1}}{4\delta_{2}}\pm\sqrt{\left(
\frac{\delta_{1}}{4\delta_{2}}\right)  ^{2}+\frac{1}{2}}. \label{20c}%
\end{equation}

In the case when $|\delta_{1}/\delta_{2}|>1$ the only physically sensible
solutions are given by the equation%
\begin{equation}
\cos2\widehat{\varphi}_{0}=\frac{\delta_{1}}{4\delta_{2}}\left[
\sqrt{1+8\left(  \frac{\delta_{2}}{\delta_{1}}\right)  ^{2}}-1\right]  .
\label{20d}%
\end{equation}

When $|\delta_{1}/\delta_{2}|<1$ the solutions corresponding to both signs in
Eq. (\ref{20c}) can be realized. This is the hysteresis domain: the type of
solution here depends on the prehistory of magnetic field variation. At
special points $H_{n}=\pi\hbar cn/\left(  eL\ell_{\mathrm{eff}}\right)  $,
where $L=\pi nL_{H}$ \ ($n=0,1,2...$), both states on each period start to be equivalent.

Let us return to Eq. (\ref{iphitot}). It can be rewritten by means of the
functions $\delta_{1}\left(  H\right)  $ and $\delta_{2}\left(  H\right)  $
(see Eq. (\ref{delty}) as%
\begin{equation}
\delta_{1}\sin2\varphi_{0}+\delta_{2}\sin4\varphi_{0}=\frac{I}{j_{c}S}.
\label{20e}%
\end{equation}

This equation can be solved exactly in the algebraic functions, but for
simplicity we will find its solutions, $\varphi_{0}^{\mathrm{\min}}$ and
$\varphi_{0}^{\mathrm{sdl}}$ in the framework of the perturbation theory under
the assumption \textbf{ }$|\delta_{1}|\gg|\delta_{2}|.$ Simple algebra leads
to the result:%
\begin{align}
&  \frac{1}{2}\arcsin\left(  \frac{I}{j_{c}S\delta_{1}}\right)  -\frac
{\delta_{2}}{\delta_{1}}\left(  \frac{I}{j_{c}S\delta_{1}}\right) \nonumber\\
&  =\left\{
\begin{tabular}
[c]{c}%
$\varphi_{0}^{\mathrm{\min}},$ if $I/j_{c}S\delta_{1}>0$\\
$\varphi_{0}^{\mathrm{sdl}},$ if $I/j_{c}S\delta_{1}<0$%
\end{tabular}
\ \ \ \ \ \ \right.  \label{20f}%
\end{align}
and%
\begin{align}
&  \frac{\pi}{2}\mathrm{sign}\left(  \frac{I}{j_{c}S\delta_{1}}\right)
-\frac{1}{2}\arcsin\left(  \frac{I}{j_{c}S\delta_{1}}\right) \nonumber\\
&  -\frac{\delta_{2}}{\delta_{1}}\left(  \frac{I}{j_{c}S\delta_{1}}\right)
=\left\{
\begin{tabular}
[c]{c}%
$\varphi_{0}^{\mathrm{sdl}},$ if $I/j_{c}S\delta_{1}>0$\\
$\varphi_{0}^{\mathrm{\min}},$ if $I/j_{c}S\delta_{1}<0$%
\end{tabular}
\ \ \ \ \ \ \right.  . \label{20g}%
\end{align}

The Eqs. (\ref{20f})-(\ref{20g}) were obtained with the help of Eq.
(\ref{20**}). Inserting the values of $\varphi_{0}^{\mathrm{\min}}$ and
$\varphi_{0}^{\mathrm{sdl}}$ to Eq. (\ref{aphi0}) one can obtain the final
expression for the difference of actions on the extremal trajectories
\begin{align}
\Delta A_{\min}^{\mathrm{sdl}}  &  =A\left[  \varphi_{0}^{\mathrm{sdl}%
}\right]  -A\left[  \varphi_{0}^{\min}\right] \nonumber\\
&  =-\frac{\hbar I}{eT}\left[  \pi/2-\arcsin\left\vert \frac{I}{j_{c}%
S\delta_{1}}\right\vert \right] \nonumber\\
&  +\frac{\hbar j_{c}S}{eT}\sqrt{1-\left(  \frac{I}{j_{c}S\delta_{1}}\right)
^{2}}\left(  \frac{L_{H}}{L}\right)  \sin\left(  \frac{L}{L_{H}}\right)
\cdot\nonumber\\
&  \left\{  1+2\left(  \frac{\delta_{2}}{\delta_{1}}\frac{I}{j_{c}S\delta_{1}%
}\right)  ^{2}\right\}  \mathrm{sign}\left(  \frac{I}{j_{c}S\delta_{1}%
}\right)  , \label{deltaA}%
\end{align}
which determine in explicit form the exponential factor of the escape time
(\ref{escdeltaA}). Looking at it one can notice the nontrivial oscillatory
type dependence of the escape time on the value of external magnetic field,
which we will discuss in the final part of the article.

\section{The value of effective action at trajectories close to the extremal
(pre-exponential factor)}

The expression (\ref{aphi0})\ obtained above allows to determine the escape
rate with the exponential accuracy, what indeed was done above. Determination
of the pre-exponential factor is more delicate task which requires knowledge
of the shape of trajectories close to the extremal one with further functional
integration of the action over them. Now we pass to perform this program.

In real situation the bifurcation point $T_{0}$ lies always much below the
critical temperature: $T_{0}\ll T_{c}.$For temperatures $T_{0}<T\ll T_{c},$ or
$|T_{0}-T|\ll T_{c}$ the general expression (\ref{Aphi}) can be considerably
simplified:\begin{widetext}
\begin{align}
A\left[  \varphi\left(  x,\tau\right)  \right]   &  =\frac{1}{L}\int
_{-1/2T}^{1/2T}d\tau\int_{-L/2}^{L/2}dx\left\{  \frac{C}{2e^{2}}\left(
\frac{\partial\varphi\left(  x,\tau\right)  }{\partial\tau}\right)  ^{2}%
-\frac{\hbar}{e}I\varphi\left(  x,\tau\right)  \right.  -\frac{\hbar j_{c}%
S}{2e}\cos\left[  2\varphi\left(  x,\tau\right)  \right] \nonumber\\
&  +\left.  \frac{\hbar}{4\pi R_{\mathrm{sh}}e^{2}}\int_{-1/2T}^{1/2T}%
d\tau_{1}\left[  \frac{\varphi\left(  x,\tau\right)  -\varphi\left(
x,\tau\right)  }{\tau-\tau_{1}}\right]  ^{2}+\frac{\hbar j_{c}S}{2e}%
\lambda_{J}^{2}\left(  \frac{\partial\varphi\left(  x,\tau\right)  }{\partial
x}-\frac{e\ell_{\mathrm{eff}}H_{\mathrm{ext}}^{2}}{\hbar c}\right)
^{2}\right\}  . \label{acshort}%
\end{align}
\end{widetext}

Let us find the solutions of Eq. (\ref{expan1}) in the vicinity of the both
time independent extremal trajectories\ (surfaces). We will look for them in
the form:%
\begin{equation}
\varphi_{n}^{k}\left(  x,\tau\right)  =\sqrt{T}\exp\left(  i2\pi
Tn\tau\right)  \chi_{n}^{k}\left(  x\right)  . \label{coshi}%
\end{equation}
Substitution of this expression in Eqs. ( \ref{Aphi})-(\ref{expan1}) leads to
equations for $\chi_{n}^{k}\left(  x\right)  :$%

\begin{align}
&  -\lambda_{J}^{2}\frac{\partial^{2}\chi_{n}^{k}\left(  x\right)  }{\partial
x^{2}}+\left[  \zeta\left(  n^{2}+Q_{\mathrm{sh}}|n|\right)  \right.
\nonumber\\
&  \left.  +\cos\left(  2\varphi^{\mathrm{extr}}\left(  x\right)  \right)
\right]  \chi_{n}^{k}\left(  x\right)  =\frac{e}{2\hbar j_{c}}\Lambda_{n}%
^{k}\chi_{n}^{k}\left(  x\right)  , \label{eigen1}%
\end{align}
where we have introduced the Q-factor $Q_{\mathrm{sh}}^{-1}=2\pi
TR_{\mathrm{sh}}C^{\ast}/\hbar\ $and the parameter $\zeta=2\pi^{2}T^{2}%
C^{\ast}/\left(  e\hbar j_{c}S\right)  .$

Note, that at the point $T_{0}$\ one has%
\begin{equation}
\Lambda_{\pm1}^{0}\left(  T_{0}\right)  =0, \label{lambda0}%
\end{equation}
since namely at this point appears the first time-dependent solution for the
extremal trajectory.

Let us recall that Eq. (\ref{eigen1}) \ is valid for both extremal
trajectories $\varphi^{\mathrm{sdl}}$ and $\varphi^{\min}$. We will look for
its solutions in the form of perturbation theory series in parameter $\left(
L/\lambda_{J}\right)  ^{2}$. There are two sets of corresponding
eigenfunctions. In the zero order approximation they can be odd or even in $x$.

For even values of $k$ $\ (k=2N,N=0,1,2...)$ we have%

\begin{align}
\chi_{\pm n}^{2N}\left(  x\right)  =\cos\left(  \frac{2\pi N}{L}x\right)
+\gamma_{\pm n}^{2N}\left(  x\right)  ,\qquad\nonumber\\
\int_{-L/2}^{L/2}dx\cos\left(  \frac{2\pi N}{L}x\right)  \gamma_{\pm n}%
^{2N}\left(  x\right)  =0. \label{eq1}%
\end{align}

For odd values of $k$ $\ (k=2N+1,N=0,1,2...)$ the eigenfunctions have form%

\begin{align}
\chi_{\pm n}^{2N+1}\left(  x\right)  =\sin\left(  \frac{\pi\left(
2N+1\right)  }{L}x\right)  +\gamma_{\pm n}^{2N+1}\left(  x\right)
,\qquad\nonumber\\
\int_{-L/2}^{L/2}dx\sin\left(  \frac{\pi\left(  2N+1\right)  }{L}x\right)
\gamma_{\pm n}^{2N+1}\left(  x\right)  =0. \label{eq2}%
\end{align}

The last integrals in Eqs. (\ref{eq1}) and (\ref{eq2}) express the relations
of orthogonality between the first order correction $\gamma_{\pm n}^{k}\left(
x\right)  $ and corresponding zero approximation solution. The functions
$\gamma$ are supposed to be small enough: $|\gamma_{\pm n}^{k}\left(
x\right)  |\ll1.$

For $k\neq0$ one can restrict consideration by the main approximation only and
get from Eqs. (\ref{lambda0}),(\ref{eq1}),(\ref{eq2}) the following expression
for the eigenvalues $\Lambda_{\pm n}^{k}$
\begin{align}
\frac{e}{2\hbar j_{c}}\Lambda_{\pm n}^{k}  &  =\lambda_{J}^{2}\left(
\frac{\pi k}{L}\right)  ^{2}+\zeta\left(  n^{2}+Q_{\mathrm{sh}}|n|\right)
\nonumber\\
&  +\frac{L_{H}}{L}\sin\left(  \frac{L}{L_{H}}\right)  \left[  1+\frac
{1}{1-\left(  \frac{L_{H}}{L}\pi k\right)  ^{2}}\right]  \cos\left(
2\varphi_{0}\right)  . \label{main}%
\end{align}
For the eigenvalues with $k=0$ ($\Lambda_{\pm n}^{0})$ we have to find the
eigenvalues up to the first correction term in parameter $\left(
L/\lambda_{J}\right)  ^{2}$. From Eqs. (\ref{lambda0}),(\ref{eq1}) follows the
equation for $\gamma_{\pm n}^{0}\left(  x\right)  :$%
\begin{align}
-\lambda_{J}^{2}\frac{\partial^{2}\gamma_{\pm n}^{0}\left(  x\right)
}{\partial x^{2}}+\cos\left(  2\varphi_{0}-\frac{2x}{L_{H}}\right) \nonumber\\
-\frac{L_{H}}{L}\sin\left(  \frac{L}{L_{H}}\right)  \cos2\varphi_{0} =0.
\label{main1}%
\end{align}
Its solution, in view of the boundary condition Eq. (\ref{bound}), is:
\begin{align}
\gamma_{\pm n}^{0}\left(  x\right)   &  =\left(  \frac{L_{H}}{2\lambda_{J}%
}\right)  ^{2}\left[  \frac{2x}{L_{H}}\cos\left(  \frac{L}{L_{H}}\right)
-\sin\left(  \frac{2x}{L_{H}}\right)  \right]  \sin2\varphi_{0}\nonumber\\
&  -\left(  \frac{L_{H}}{2\lambda_{J}}\right)  ^{2}\left[  \cos\left(
\frac{2x}{L_{H}}\right)  -\frac{L_{H}}{L}\sin\left(  \frac{L}{L_{H}}\right)
\right. \nonumber\\
&  \left.  +\frac{L_{H}}{L}\sin\left(  \frac{L}{L_{H}}\right)  \left(
\frac{2x^{2}}{L_{H}^{2}}-\frac{L^{2}}{6L_{H}^{2}}\right)  \right]
\cos2\varphi_{0}. \label{gamma0}%
\end{align}
>From Eqs. (\ref{eigen1}),(\ref{eq1}),(\ref{gamma0}) one finds the value of
$\Lambda_{\pm n}^{0}$ with the first correction terms (compare to Eq.
(\ref{main}) ) :%
\begin{align}
\frac{e}{2\hbar j_{c}}\Lambda_{\pm n}^{0}  &  =\zeta\left(  n^{2}%
+Q_{\mathrm{sh}}|n|\right)  +\frac{L_{H}}{L}\sin\left(  \frac{L}{L_{H}%
}\right)  \cos2\varphi_{0}\nonumber\\
&  +\left(  \frac{L_{H}}{2\lambda_{J}}\right)  ^{2}\kappa\left(  \frac{L_{H}%
}{L}\right)  \cos4\varphi_{0}. \label{mainapr}%
\end{align}
One can notice, that the eigenvalue $\Lambda_{0}^{0}$ in the vicinity of the
saddle point trajectory is negative. This property is an obvious consequence
of the Eqs. (\ref{20**})-(\ref{sico}) and namely this fact results in the
appearance of the imaginary part of the partition function (\ref{PF}).

Substitution of $n=1$ to Eq. (\ref{mainapr}) gives us the explicit definition
of the crossover temperature $T_{0}$\cite{OBV07}$:$%
\begin{align}
\zeta\left(  T_{0}\right)  \left(  1+Q_{\mathrm{sh}}\left(  T_{0}\right)
\right)  +\frac{L_{H}}{L}\sin\left(  \frac{L}{L_{H}}\right)  \cos2\varphi
_{0}\nonumber\\
+\left(  \frac{L_{H}}{2\lambda_{J}}\right)  ^{2}\kappa\left(  \frac{L_{H}}%
{L}\right)  \cos4\varphi_{0} =0\text{ ,} \label{43}%
\end{align}
where $\varphi_{0}=$ $\varphi^{\mathrm{sdl}}$ and it is the function of
external current $I$\ and magnetic field $H$ (see Eqs. (\ref{iphitot})-
(\ref{20f})). For the critical value of current, where $\partial
I/\partial\varphi_{0}=0,$ $T_{0}=0.$ This equality follows immediately from
equations Eq. (\ref{43}). Note that our parameter of perturbation theory is
$\left(  L_{H}/\lambda_{J}\right)  ^{2}$ \ and the corrections to
eigenfunctions (Eqs. (\ref{eq1}) and (\ref{gamma0})) are small by this
parameter for all values of the external current $I$ and magnetic field $H$.
That\textbf{ }means that the Eq. (\ref{iphitot}) is valid even in the vicinity
of the points, where $\sin\left(  L/L_{H}\right)  =0.$ Important that in these
regions both terms in the r.h.s. of the Eq. (\ref{iphitot}) are of the same
order and the nontrivial dependence of $\varphi_{0}\left(  I,H\right)  $
arises, since Eq. (\ref{iphitot}) is equivalent to the fourth order equation
and in considered region all its coefficients turn out to be of the same order.

Now one can write down the expression for the effective action (\ref{acshort}%
), \ valid near both extremal trajectories. It is enough to substitute in Eq.
(\ref{acshort}) function $\varphi\left(  x,\tau\right)  $ in the form of Eq.
(\ref{expan}) with $\varphi_{n}^{k}\left(  x,\tau\right)  $ defined by Eqs.
(\ref{coshi}),(\ref{eq1}) and (\ref{gamma0})). \ Since the eigenvalues
$\Lambda_{\pm1}^{0}$ tend to zero as $T\rightarrow T_{0}$ for saddle point
trajectory we have to keep in expansion of the effective action over the
coefficients $\{B_{\pm1}^{0}\}$ all terms up to the fourth order, keeping also
the products of the type\ \{$\left(  B_{1}^{0}\right)  ^{2}B_{-2}^{k},\left(
B_{-1}^{0}\right)  ^{2}B_{2}^{k}$\}. In result of integration over $x$ and
$\tau$ (see Appendix A) the action (\ref{acshort}) takes the explicit form as
the function of coefficients $B_{n}^{k}$:%
\begin{align}
&  A\left[  B_{0}^{0},B_{1}^{0},B_{-1}^{0},..B_{n}^{k}..\right]  =A\left[
\varphi_{\mathrm{extr}}\left(  x\right)  \right] \nonumber\\
&  -\gamma_{1}\left[  2B_{1}^{0}B_{-1}^{0}B_{0}^{0}+\left(  B_{1}^{0}\right)
^{2}B_{-2}^{0}+\left(  B_{-1}^{0}\right)  ^{2}B_{2}^{0}\right] \nonumber\\
&  -\gamma_{2}\left(  B_{1}^{0}\right)  ^{2}\left(  B_{-1}^{0}\right)
^{2}+\frac{1}{2}\sum_{n,k}\Lambda_{n}^{k}|B_{n}^{k}|^{2}. \label{Afinal1}%
\end{align}
Calculation of the coefficients $\gamma_{1,2}$ is cumbersome but
straightforward. It is necessary to remember that the functions $\chi_{\pm
n}^{k}\left(  x\right)  $ before to be used in Eq. (\ref{acshort}) should be
normalized. In result one finds%

\begin{align}
\gamma_{1}  &  =\frac{2\hbar j_{c}}{e}\left(  \frac{T}{S}\right)
^{1/2}\left\{  \frac{L_{H}}{L}\sin\left(  \frac{L}{L_{H}}\right)  \sin
2\varphi_{0}\right.  +\nonumber\\
&  \left.  2\left(  \frac{L_{H}}{2\lambda_{J}}\right)  ^{2}\kappa\left(
\frac{L_{H}}{L}\right)  \sin4\varphi_{0}\right\}  , \label{gamma1}%
\end{align}%
\begin{align}
\gamma_{2}  &  =\frac{2\hbar j_{c}}{e}\left(  \frac{T}{S}\right)  \left\{
\frac{L_{H}}{L}\sin\left(  \frac{L}{L_{H}}\right)  \cos2\varphi_{0}+\right.
\nonumber\\
&  \left.  \frac{1}{2}\left(  \frac{L_{H}}{2\lambda_{J}}\right)  ^{2}\left[
5\kappa\left(  \frac{L_{H}}{L}\right)  \cos4\varphi_{0}+3\mu\left(
\frac{L_{H}}{L}\right)  \right]  \right\}  . \label{gamma2}%
\end{align}

\section{Oscillations of the Escape Time vs Magnetic Field}

Now we are ready to calculate the escape time of the "small" JJ, which is
given by\ Eq. (\ref{escape}). The imaginary part of the partition function
$\operatorname{Im}Z$ in Eq. (\ref{escape}) is determined by the integral over
trajectories close to the saddle point trajectory $\varphi^{\mathrm{sdl}%
}\left(  x\right)  $ (see Eq. (\ref{expan}) ). Corresponding expression for
action was already obtained above and it is given by Eq. (\ref{Afinal1}). The
functional integral in Eq. (\ref{escape}) is reduced now to integration over
all coefficients $B_{n}^{k}.$

Let us start from the integration over the coefficient $B_{0}^{0}.$ Since the
eigenvalue $\Lambda_{0}^{0}$ in vicinity of the saddle point trajectory is
negative it requires special considerations. In order to get the finite answer
one has before integration to perform analytical continuation $B_{0}%
^{0}\rightarrow ib$ and only after that to carry out the integral:%
\begin{align}
&  \int_{-\infty}^{\infty}\frac{dB_{0}^{0}}{\sqrt{2\pi}}\exp\left[  -\frac
{1}{2}\Lambda_{0}^{0}\left(  B_{0}^{0}\right)  ^{2}+2\gamma_{1}B_{0}%
^{0}\left(  B_{1}^{0}B_{-1}^{0}\right)  \right] \nonumber\\
&  \rightarrow i\int_{-\infty}^{\infty}\frac{db}{\sqrt{2\pi}}\exp\left[
-\frac{1}{2}|\Lambda_{0}^{0}|b^{2}+2i\gamma_{1}b\left(  B_{1}^{0}B_{-1}%
^{0}\right)  \right] \nonumber\\
&  =\frac{i}{\sqrt{|\Lambda_{0}^{0}|}}\exp\left(  -\frac{2\gamma_{1}^{2}%
}{|\Lambda_{0}^{0}|}\left(  B_{1}^{0}B_{-1}^{0}\right)  ^{2}\right)  .
\label{integ1}%
\end{align}

Integration of the coefficients $B_{\pm1}^{0},$ we will yet leave for further
consideration and now perform that one over the coefficients $B_{\pm2}^{0}$ in
accordance with the formula
\begin{align}
\int_{-\infty}^{\infty}\frac{d^{2}B_{\pm2}^{0}}{2\pi}\exp\left\{  -\Lambda
_{2}^{0}\left(  B_{2}^{0}\right)  ^{2}+\gamma_{1}\left[  \left(  B_{1}%
^{0}\right)  ^{2}B_{-2}^{0}\right.  \right. \nonumber\\
\left.  \left.  +\left(  B_{-1}^{0}\right)  ^{2}B_{2}^{0}\right]  \right\}
=\frac{1}{2\Lambda_{2}^{0}}\exp\left(  \frac{\gamma_{1}^{2}}{\Lambda_{2}^{0}%
}\left(  B_{1}^{0}B_{-1}^{0}\right)  ^{2}\right)  . \label{integ2}%
\end{align}

Integrations over the remaining coefficients, besides are of the canonical
Gaussian type and can be easily performed.

What concerns calculation of the real part of the partition function
$\operatorname{Re}Z$ it is determined by integration over the trajectories
passing close to the minimal trajectory $\varphi^{\mathrm{\min}}\left(
x\right)  $ and\ in this case one can take the action (\ref{Afinal1}) only
with quadratic accuracy over $B_{n}^{k}.$

Performing discussed above integrations in real and imaginary parts of the
partition function (see Ref. \cite{LO84})) one finds the escape time for high
enough temperatures $T>T_{0}$, or in the narrow vicinity of $T_{0}$
($|T-T_{0}|\ll T_{0}$): \begin{widetext}
\begin{eqnarray}
&\tau_{\mathrm{esc}}^{-1}= 2T_{0} \exp\left\{  -\Delta A_{\min}^{\mathrm{sdl}%
}\right\} \cdot \nonumber \\
& \left\{ \frac{1}{2\sqrt{|\Lambda_{0}^{0}|}}\int_{-\infty}^{\infty
}\frac{d^{2}B_{1}^{0}}{2\pi}\exp\left[  -\Lambda_{1}^{0}|B_{1}^{0}|^{2}%
-|B_{1}^{0}|^{4}\left(  -\gamma_{2}+\gamma_{1}^{2}\left(  \frac{2}%
{|\Lambda_{0}^{0}|}-\frac{1}{\Lambda_{2}^{0}}\right)  \right)  \right]
\right\}_{\mathrm{sdl}}\left\{\sqrt{|\Lambda_{0}^{0}|}\right\}_{\min}Y_{1}Y_{2}, \label{esc1}%
\end{eqnarray}
with
\begin{equation}
Y_{1}=\left.  \left\{ {\displaystyle\prod\limits_{k=1}^{\infty}}
\frac{1}{\sqrt{\Lambda_{0}^{k}}}\left(
{\displaystyle\prod\limits_{n=1}^{\infty}}
\frac{1}{\left(  2\Lambda_{n}^{k}\right)  }\right)  \right\}  _{\mathrm{sdl}%
}\right/  \left\{ {\displaystyle\prod\limits_{k=1}^{\infty}}
\frac{1}{\sqrt{\Lambda_{0}^{k}}}
{\displaystyle\prod\limits_{n=1}^{\infty}}
\frac{1}{\left(  2\Lambda_{n}^{k}\right)  }\right\}  _{\min}, \label{Y1}%
\end{equation}
\begin{equation}
Y_{2}=\left.  \left\{ {\displaystyle\prod\limits_{n=2}^{\infty}}
\frac{1}{\left(  2\Lambda_{n}^{0}\right)  }\right\}
_{\mathrm{sdl}}\right/ \left\{
{\displaystyle\prod\limits_{n=1}^{\infty}}
\frac{1}{\left(  2\Lambda_{n}^{0}\right)  }\right\}  _{\min}. \label{Y2}%
\end{equation}
\end{widetext}
Note that the pre-factor in Eq. (\ref{esc1}) contains $T_{0}$ instead of $T$.
The eigenvalues $\{\Lambda_{k}^{n}\}_{\left[  \min,\mathrm{sdl}\right]  }$ are
defined by Eqs. (\ref{main}), (\ref{mainapr}).

The remaining integral over $B_{1}^{0}$ \ in Eq. (\ref{esc1}) \ can be
expressed in terms of Fresnel integral%

\begin{equation}
\Phi\left(  x\right)  =\frac{2}{\sqrt{\pi}}\int_{0}^{x}dt\exp\left(
-t^{2}\right)
\end{equation}
and one obtains $\tau_{\mathrm{esc}}^{-1}$ in the final form:%

\begin{align}
&  \tau_{\mathrm{esc}}^{-1}=\frac{1}{4}\sqrt{\frac{\pi}{B}}T_{0}\exp\left(
-\Delta A_{\min}^{\mathrm{sdl}}\right)  Y_{1}Y_{2} \left\{  \sqrt{\Lambda
_{0}^{0}}\right\}  _{\min}\dot\nonumber\\
&  \left\{  \frac{1}{\sqrt{|\Lambda_{0}^{0}|}}\left[  1-\Phi\left(
\frac{\Lambda_{1}^{0}}{2\sqrt{B}}\right)  \right]  \exp\frac{\left(
\Lambda_{1}^{0}\right)  ^{2}}{4B}\right\}  _{\mathrm{sdl}}, \label{42}%
\end{align}
with
\begin{equation}
B=-\gamma_{2}+\gamma_{1}^{2}\left(  \frac{2}{|\Lambda_{0}^{0}|}-\frac
{1}{\Lambda_{2}^{0}}\right)  _{\mathrm{sdl}}.
\end{equation}

Using the explicit Eq. (\ref{mainapr}) for eigenfunctions $\Lambda_{n}^{0}$
one can present Eq. (\ref{Y2}) in terms of Euler gamma-function $\Gamma\left(
x\right)  $:
\[
Y_{2}=\frac{4\hbar j_{c}\zeta}{e}\frac{\left[  \Gamma\left(  2-n_{1}\left(
H\right)  \right)  \Gamma\left(  2-n_{2}\left(  H\right)  \right)  \right]
_{\mathrm{sdl}}}{\left[  \Gamma\left(  1-n_{1}\left(  H\right)  \right)
\Gamma\left(  1-n_{2}\left(  H\right)  \right)  \right]  _{\min}},
\]
while the values $\left[  n_{1,2}\right]  _{\mathrm{saddle}}$,$\left[
n_{1,2}\right]  _{\min}$ are roots of equation%

\begin{align}
\zeta\left(  T\right)  \left[  n^{2}+Q_{\mathrm{sh}}\left(  T\right)
n\right]  +\frac{L_{H}}{L}\sin\left(  \frac{L}{L_{H}}\right)  \cos2\varphi
_{0}\nonumber\\
\text{ }+\left(  \frac{L_{H}}{2\lambda_{J}}\right)  ^{2}\kappa\left(
\frac{L_{H}}{L}\right)  \cos4\varphi_{0} =0\text{,} \label{roots}%
\end{align}
written for $\varphi_{0}=\left\{  \varphi_{0}^{\mathrm{saddle}},\varphi
_{0}^{\min}\right\}  $ accordingly. From Eq. (\ref{roots}) one can see that
the finiteness of $L$ leads to strong variation of $Y_{2}$ and, consequently
$\tau_{\mathrm{esc}}^{-1},$ as the function of magnetic field even for small
junction with $L\ll\lambda_{J}.$

\begin{figure}[b]
\includegraphics[width=1.0\columnwidth]{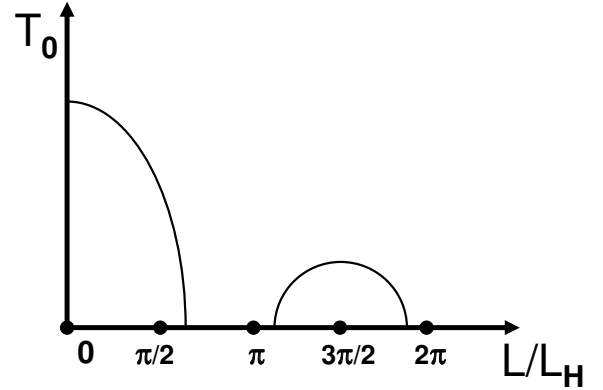}\caption{Schematic dependence
of crossover temperature on magnetic field}%
\label{t0}%
\end{figure}

All quantities $\Delta A_{\min}^{\mathrm{sdl}},T_{0},$ $\tau_{\mathrm{esc}%
}^{-1}$ are oscillatory functions versus magnetic field (see Eqs.(\ref{Aphi}),
(\ref{iphitot}),(\ref{aphi0}),(\ref{deltaA}),(\ref{43}),(\ref{Y1})). As the
example let us consider the behavior of $T_{0}(H)$. In first approximation by
parameter $\left(  L/\lambda_{J}\right)  ^{2}$ one can obtain from
Eqs.(\ref{20e})-(\ref{43}) following simple expression for the crossover
temperature $T_{0}\left(  H\right)  :$%
\begin{align*}
T_{0}^{2}+\varepsilon T_{0}-\frac{ej_{c}SR_{\mathrm{sh}}}{\pi}%
\varepsilon\sqrt{\delta_{1}^{2}\left(  H\right)  -\left(  \frac{I}{j_{c}%
S}\right)  ^{2}}\\
-2\frac{\delta_{2}\left(  H\right)  }{\delta_{1}^{2}\left(  H\right)  }\left(
\delta_{1}^{2}\left(  H\right)  -\left(  \frac{I}{j_{c}S}\right)  ^{2}\right)
 =0
\end{align*}
with $\varepsilon=Q_{\mathrm{sh}}T=\hbar/\left(  2\pi R_{\mathrm{sh}}C^{\ast
}\right)  \ ,$ $\delta_{1}$ and $\delta_{2}$ defined by Eq. (\ref{delty}%
).\ Its physical solution \begin{widetext}
\begin{equation}
T_{0}\left(  H\right)  =\frac{\varepsilon}{2}\left\{  \left\{
1+\frac {4ej_{c}SR_{\mathrm{sh}}}{\pi\varepsilon}\left[
\sqrt{\delta_{1}^{2}\left( H\right)  -\left( \frac{I}{j_{c}S}\right)
^{2}}-2\frac{\delta_{2}\left( H\right) }{\delta_{1}^{2}\left(
H\right)  }\left(  \delta_{1}^{2}\left( H\right)  -\left(
\frac{I}{j_{c}S}\right)  ^{2}\right)  \right] \right\}
^{1/2}-1\right\}  \label{t0sol}%
\end{equation}
\end{widetext}evidently oscillates versus magnetic field as it is sketched in
Fig. \ref{t0}. Eq. (\ref{t0sol}) can be used until the second term in square
parenthesis is smaller than the first one. Close to the special points ($L=\pi
nL_{H}$) this condition can be not valid more and in such a case
Eqs.(\ref{iphitot}),(\ref{roots}) should be solved exactly. 

In contrast to $Y_{2},$ the pre-factor $Y_{1}$ contains only terms with
$\Lambda_{0}^{k}$ ($k\neq0)$ and in result depends on magnetic field weakly.
Nevertheless this dependence turns out to be singular due to the logarithmic
divergence of the product in Eq. (\ref{Y1}). Details are presented in Appendix B.

Eq. (\ref{42}) enable us to estimate the width of the crossover region between
Arrenius and macroscopic quantum tunneling regimes. It can be found from
condition that the argument of the Fresnel function is of the order of one:%
\begin{equation}
\left\{  \Lambda_{1}^{0}\right\}  _{\mathrm{sdl}}\sim2\sqrt{B}. \label{41}%
\end{equation}

\section{Final remarks}

Even a weak external magnetic field can change strongly the value of the
critical current of the Josephson junction of a small size. Note, that some
special points of external magnetic field appear in the problem under
consideration. In the vicinity of these points \ the perturbation theory fails
and the escape time can be calculated only by means of the exact solution of
the equations on $T_{0}.$ and $\varphi_{0}.$ It worth to mention that our
general equations enable us to consider even such points.

In the vicinity of the crossover temperature $T_{0}$ from Arrenius law to the
quantum tunneling, one can observe the strong effect of the finiteness of the
junction length $L$ even in the pre-exponential factor.

\section{Acknowledgements}

We thank A.~Ustinov and K.~Fedorov for useful discussions. The work was
partially supported by MIUR under the project PRIN 2006 "Effetti quantistici
macroscopici e dispositivi superconduttivi" and the EU Project MIDAS
"Macroscopic Interference Devices for Atomic Solid State Systems". Yu.N.O. and
A.A.V acknowledges support of the grants RFBR-07-02-12058 and RFBR-06-02-16223.

\section{Appendix A}

The eigenvalues $\Lambda_{\pm1}^{0}$ at the saddle point trajectory tend to
zero as $T\rightarrow T_{0}.$ In result, when one substitutes the function
$\varphi\left(  \mathbf{r},\tau\right)  $ in the form Eq. (\ref{expan}) \ to
the action Eq. (\ref{acshort})\ and then expands the expression for action in
Taylor series he has to keep "dangerous"\ terms, containing \ $B_{\pm1}^{0}$,
up to the fourth order. Moreover one has to keep also products of the type
\{$\left(  B_{1}^{0}\right)  ^{2}B_{-2}^{k},\left(  B_{-1}^{0}\right)
^{2}B_{2}^{k}$\}. All other terms (containing $B_{n}^{k}$ with $k\neq0$\ and
$n\neq0,\pm1)$\ is enough to take in the second order approximation. In this
way one writes the value of the effective action (\ref{acshort}) \ valid near
both extremal trajectories in the form (see Ref. \cite{OBV07})
\begin{widetext}
\begin{align}
A\left[  \varphi\left(  x,\tau\right)  \right]   &  =A\left[
\varphi _{\mathrm{extr}}\left(  x\right)  \right]
+\frac{1}{2}\sum_{n,k}\Lambda
_{n}^{k}|B_{n}^{k}|^{2}-\frac{2\hbar Sj_{c}}{e}\sqrt{T}\int_{-L/2}^{L/2}%
\frac{dx}{L}\sin\left(  2\varphi_{\mathrm{extr}}\left(  x\right)
\right)
\nonumber\\
&  \left\{
2\frac{B_{1}^{0}B_{-1}^{0}\chi_{1}^{0}\chi_{-1}^{0}}{||\chi
_{1}^{0}||||\chi_{-1}^{0}||}\sum_{k}\frac{B_{0}^{k}\chi_{0}^{k}}{||\chi
_{0}^{k}||}+\frac{\left(  B_{1}^{0}\right)  ^{2}\left(
\chi_{1}^{0}\right)
^{2}}{||\chi_{1}^{0}||^{2}}\sum_{k}\frac{B_{-2}^{k}\chi_{-2}^{k}}{||\chi
_{-2}^{k}||}+\frac{\left(  B_{-1}^{0}\right)  ^{2}\left(  \chi_{-1}%
^{0}\right)  ^{2}}{||\chi_{-1}^{0}||^{2}}\sum_{k}\frac{B_{2}^{k}\chi_{2}^{k}%
}{||\chi_{2}^{k}||}\right\} \nonumber\\
&  -\frac{2\hbar Sj_{c}}{e}T\int_{-L/2}^{L/2}\frac{dx}{L}\cos\left[
2\varphi_{\mathrm{extr}}\left(  x,\tau\right)  \right]  \frac{\left(
B_{1}^{0}\right)  ^{2}\left(  B_{-1}^{0}\right)  ^{2}}{||\chi_{1}^{0}%
||^{2}||\chi_{-1}^{0}||^{2}}\left(  \chi_{1}^{0}\right)  ^{2}\left(
\chi
_{-1}^{0}\right)  ^{2}, \label{action32}%
\end{align}
\end{widetext}where $||..||$ is the norm of function. Let us carry out the
integrals entering in the Eq. (\ref{action32}) in the explicit form:%
\begin{align}
I_{1}  &  =\int_{-L/2}^{L/2}\frac{dx}{L}\sin\left(  2\varphi_{\mathrm{extr}%
}\left(  x\right)  \right)  \frac{\chi_{1}^{0}\chi_{-1}^{0}\chi_{0}^{0}%
}{||\chi_{1}^{0}||||\chi_{-1}^{0}||||\chi_{0}^{0}||}\nonumber\\
&  =\frac{1}{S^{3/2}}\left[  \frac{L_{H}}{L}\sin\left(  \frac{L}{L_{H}%
}\right)  \sin2\varphi_{0}\right. \nonumber\\
&  +\left.  2\left(  \frac{L_{H}}{2\lambda_{J}}\right)  ^{2}\kappa\left(
\frac{L_{H}}{L}\right)  \sin4\varphi_{0}\right]  \label{int1}%
\end{align}
and
\begin{align}
I_{2}  &  =\int_{-L/2}^{L/2}\frac{dx}{L}\cos\left(  2\varphi_{\mathrm{extr}%
}\left(  x\right)  \right)  \frac{\left(  \chi_{1}^{0}\right)  ^{2}\left(
\chi_{-1}^{0}\right)  ^{2}}{||\chi_{1}^{0}||^{2}||\chi_{-1}^{0}||^{2}%
}\nonumber\\
&  =\frac{1}{S^{2}}\left\{  \frac{L_{H}}{L}\sin\left(  \frac{L}{L_{H}}\right)
\cos2\varphi_{0}\right. \nonumber\\
&  +\left.  \frac{1}{2}\left(  \frac{L_{H}}{2\lambda_{J}}\right)  ^{2}\left[
5\kappa\left(  \frac{L_{H}}{L}\right)  \cos4\varphi_{0}+3\mu\left(
\frac{L_{H}}{L}\right)  \right]  \right\}  \label{int2}%
\end{align}
with the function $\mu$ defined by Eq. (\ref{mu}). Using these expressions in
Eq. (\ref{action32}) one can obtain the final expression (\ref{Afinal1}) for
the effective action, valid for trajectories close to the extremal ones.
Integrals (\ref{int1})- (\ref{int2}) \ appear in Eq. (\ref{Afinal1}) by means
of the functions\ (\ref{gamma1})- (\ref{gamma2}):%
\begin{equation}
\gamma_{1}=\frac{2\hbar j_{c}S}{e}\sqrt{T}I_{1},\qquad\gamma_{2}=\frac{2\hbar
j_{c}ST}{e}I_{2}.
\end{equation}

\section{Appendix B}

One can see that expression for the pre-exponential factor $Y_{1}$ (\ref{Y1})
is divergent at $\{n,k\}\rightarrow\infty$ (this follows from Eq.
(\ref{gamma0})). This divergency has the logarithmic character and should be
cut off at $n\sim T_{c}/T_{0}$ and $k\sim\lambda_{J}/\ell_{\mathrm{eff}}.$ As
result\ with the logarithmic accuracy one can find: \begin{widetext}
\begin{align}
Y_{1}  &  =\left(
{\displaystyle\prod\limits_{k=1}}
\frac{1}{\sqrt{\Lambda_{0}^{k}}}\left(
{\displaystyle\prod\limits_{n=1}^{\infty}}
\frac{1}{2\Lambda_{n}^{k}}\right)  \right)  _{\mathrm{sdl}}\left\{
\left(
{\displaystyle\prod\limits_{k=1}^{\infty}}
\frac{1}{\sqrt{\Lambda_{0}^{k}}}%
{\displaystyle\prod\limits_{n=1}^{\infty}}
\frac{1}{2\Lambda_{n}^{k}}\right)  \right\}  _{\min}^{-1}=\nonumber\\
&  \exp\left\{  -\frac{1}{2}\frac{L_{H}}{L}\sin\left(
\frac{L}{L_{H}}\right) \left[  \cos\left(
2\varphi_{0}^{\mathrm{sdl}}\right)  -\cos\left(
2\varphi_{0}^{\min}\right)  \right]
\sum_{k=1}^{\infty}\frac{1}{1-\left( \frac{L_{H}}{L}\pi k\right)
^{2}}\sum_{n=-\infty}^{\infty}\frac{1}{\left(
\frac{\pi k\lambda_{J}}{L}\right)  ^{2}+\zeta\left(  n^{2}+Q_{\mathrm{sn}%
}|n|\right)  }\right\}  \cdot\nonumber\\
&  \exp\left\{  -\frac{1}{2}\frac{L_{H}}{L}\sin\left(
\frac{L}{L_{H}}\right) \left[  \cos\left(
2\varphi_{0}^{\mathrm{sdl}}\right)  -\cos\left(
2\varphi_{0}^{\min}\right)  \right]  \right\}  \sum_{k=1}^{\lambda_{J}%
/\ell_{\mathrm{eff}}}\sum_{n=-T_{c}/T_{0}}^{T_{c}/T_{0}}\left[
\left(
\frac{\pi k\lambda_{J}}{L}\right)  ^{2}+\zeta\left(  n^{2}+Q_{\mathrm{sn}%
}|n|\right)  \right]  ^{-1}. \label{44}%
\end{align}
\end{widetext} The double sum in the first multiplier in the r.h.s. of this
expression converges at infinity.


\begin{references}
\bibitem{c}   The external magnetic field H is supposed to be directed along the plane of junction.
\bibitem{class} The motion in the classically allowed region is described by the same functional Eq. (7)
in the assumption $t=t_1$ (the latter eliminates third term in  Eq.
(7)).
\end{references}
\end{document}